\documentstyle[aps]{revtex}

\begin{document}
\title{Quantization of Massive Non-Abelian Gauge Fields\\
in The Hamiltonian Path Integral Formalism }
\author{}
\author{Jun-Chen Su }
\author{Center for Theoretical Physics, Department of Physics, \and Jilin
University, Changchun 130023,}
\author{People's Republic of China}
\date{}
\maketitle

\begin{abstract}
It is argued that the massive non-Abelian gauge field theory without
involving Higgs bosons may be well established on the basis of
gauge-invariance principle because the dynamics of the field is
gauge-invariant in the physical space defined by the Lorentz constraint
condition. The quantization of the field can readily be performed in the
Hamiltonian path-integral formalism and leads to a quantum theory which
shows good renormalizability and unitarity.\\PACS: 11.15.-q, 11.10.Ef
\end{abstract}

~~~~It is the prevailing viewpoint that in the non-Abelian case, a
renormalizable and unitary massive gauge field theory could not be built up
if without introducing the Higgs mechanism[1.2]. This letter will show that
this viewpoint is not always true. Originally, the Proca-type Lagrangian
density 
\begin{equation}
{\cal L}=-\frac 14F^{a\mu \nu }F_{\mu \nu }^a+\frac 12m^2A^{a\mu }A_\mu ^a
\end{equation}
was used, as a starting point, to establish the massive non-Abelian gauge
field theory. However, as was widely discussed in the literature[3-9], this
Lagrangian is not gauge-invariant and gives rise to an unrenormalizable
quantum theory. Although the Lagrangian may be recast in a gauge-invariant
Stueckelberg version [8-16], the scalar Stueckelberg function introduced in
the mass term has no physical meaning and unavoidably much complicates the
theory. In the previous studies, the Lagrangian mentioned above was
considered to form a complete description for the massive gauge field
dynamics. This concept, we note, actually is not reasonable because the
Lagrangian contains redundant unphysical degrees of freedom. As is
well-known, a massive vector field has only three polarization states. They
can completely be described by the Lorentz-covariant transverse vector
potential $A_T^{a\mu }(x)$. This vector potential defines a physical space
in which the massive gauge field exists only. According to the general
principle established well in mechanics, we should firstly write a
Lagrangian represented in terms of the independent variables $A_T^{a\mu }$
such that 
\begin{equation}
{\cal L}=-\frac 14F_T^{a\mu \nu }F_{T\mu \nu }^a+\frac 12m^2A_T^{a\mu
}A_{T\mu }^a
\end{equation}
In order to express the dynamics through the full vector potential $A_\mu ^a$
as shown in Eq.(1), it is necessary to impose an appropriate constraint
condition on the Lagrangian (1). A suitable constraint is the covariant
Lorentz condition 
\begin{equation}
\varphi ^a\equiv \partial ^\mu A_\mu ^a=0
\end{equation}
which implies vanishing of the redundant longitudinal vector potential $%
A_L^{a\mu }(x)$. It is obvious that only the Lagrangian (2), or instead, the
Lagrangian (1) plus the constraint (3) can be viewed as complete for
formulating the dynamics of the massive gauge field.

An important fact is that in the physical space, the dynamics of the massive
gauge field is gauge-invariant. In fact, under the gauge-transformation[6] 
\begin{equation}
\delta A^a_{\mu} =D^{ab}_{\mu}\theta^b
\end{equation}
where 
\begin{equation}
D^{ab}_{\mu}=\delta^{ab}\partial_{\mu}-gf^{abc}A^c_{\mu}
\end{equation}
and the condition (3), the action given by the Lagrangian (1) is invariant 
\begin{equation}
\delta S=-m^2\int d^4x\theta^a\partial^{\mu}A^a_{\mu}=0
\end{equation}
Saying equivalently, the action made of the Lagrangian (2) is invariant with
respect to the gauge-transformation as shown in Eqs.(4) and (5) with the
full vector potential being replaced by the transverse one.

The constraint (3) may be incorporated in the Lagrangian (1) by the Lagrange
undetermined multiplier method to give a generalized Lagrangian. In the
first order formalism, this Lagrangian is written as [6.18] 
\begin{equation}
{\cal L}=\frac 14F^{a\mu \nu }F_{\mu \nu }^a-\frac 12F^{a\mu \nu }(\partial
_\mu A_\nu ^a-\partial _\nu A_\mu ^a+gf^{abc}A_\mu ^bA_\nu ^c)+\frac 12%
m^2A^{a\mu }A_\mu ^a-\lambda ^a\partial ^\mu A_\mu ^a
\end{equation}
where $\lambda ^a$ are the Lagrange multipliers. Using the conjugate
variables which are defined by 
\begin{equation}
\pi _\mu ^a(x)=\frac{\partial {\cal L}}{\partial \dot A^{a\mu }}=F_{\mu
0}^a-\lambda ^a\delta _{\mu 0}=\cases{E^a_k&, $if ~\mu =k=1,2,3$;\cr
-\lambda^a=-E^a_0&, $if ~\mu=0$\cr}
\end{equation}
the Lagrangian (7) may be rewritten in the canonical form 
\begin{equation}
{\cal L}=E^{a\mu }\dot A_\mu ^a+A_0^aC^a-E_0^a\varphi ^a-{\cal H}
\end{equation}
where 
\begin{equation}
C^a=\partial ^\mu E_\mu ^a+gf^{abc}A_k^bE^{ck}+m^2A_0^a
\end{equation}
\begin{equation}
{\cal H}=\frac 12(E_k^a)^2+\frac 14(F_{ij}^a)^2+\frac 12%
m^2[(A_0^a)^2+(A_k^a)^2]
\end{equation}
here $E_\mu ^a=(E_0^a,E_k^a)$ is a Lorentz vector, ${\cal H}$ is the
Hamiltonian density. In the above, the Lorentz and the three-dimensional
spatial indices are respectively denoted by the Greek and Latin letters.
From the stationary condition of the action constructed by the Lagrangian
(7) or (9), one may derive the equations of motion as follows 
\begin{equation}
\dot A_k^a=\partial _kA_0^a+gf^{abc}A_k^bA_0^c-E_k^a
\end{equation}
\begin{equation}
\dot E_k^a=\partial
^iF_{ik}^a+gf^{abc}(E_k^bA_0^c+F_{ki}^bA^{ci})+m^2A_k^a+\partial _kE_0^a
\end{equation}
\begin{equation}
C^a(x)\equiv \partial ^\mu E_\mu ^a+gf^{abc}A_k^bE^{ck}+m^2A_0^a=0
\end{equation}
and the equation (3). Eqs.(12) and (13) are the equations of motion
satisfied by the independent canonical variables $A_k^a$ and $E_k^a$, while,
Eqs.(3) and (14) can only be regarded as the constraint equations obeyed by
the constrained variables $A_0^a$ and $E_0^a$. Eq.(9) clearly shows that
these constraints have already been incorporated in the Lagrangian by the
Lagrange undetermined multiplier method. Especially, the Lagrange
multipliers are just the constrained variables themselves in this case.

Along the general line by Dirac[17], we shall examine the evolution of the
constraints $\varphi ^a$ and $C^a$ with time. Taking the derivative of the
both equations (3) and (14) with respect to time and making use of the
equations of motion : 
\begin{equation}
\dot A_\mu ^a(x)=\frac{\delta H}{\delta E^{a\mu }(x)}-\int d^4y[A_0^b(y)%
\frac{\delta C^b(y)}{\delta E^{a\mu }(x)}-E_0^b(y)\frac{\delta \varphi ^b(y)%
}{\delta E^{a\mu }(x)}]
\end{equation}
\begin{equation}
\dot E_\mu ^a(x)=-\frac{\delta H}{\delta A^{a\mu }(x)}+\int d^4y[A_0^b(y)%
\frac{\delta C^b(y)}{\delta A^{a\mu }(x)}-E_0^b(y)\frac{\delta \varphi ^b(y)%
}{\delta A^{a\mu }(x)}]
\end{equation}
, which are obtained from the stationary condition of the action given by
the Lagrangian (9), one may derive the following consistence
equations[17,18] 
\begin{equation}
\{H,\varphi ^a(x)\}+\int d^4y[\{\varphi ^a(x),C^b(y)\}A_0^b(y)-\{\varphi
^a(x),\varphi ^b(y)\}E_0^b(y)]=0
\end{equation}
\begin{equation}
\{H,C^a(x)\}+\int d^4y[\{C^a(x),C^b(y)\}A_0^b(y)-\{C^a(x),\varphi
^b(y)\}E_0^b(y)]=0
\end{equation}
where Eqs.(3) and (14) have been used. In the above, H is the Hamiltonian
defined by an integral of the Hamiltonian density shown in Eq.(11) over the
coordinate x and $\{F,G\}$ represents the Poisson bracket which is defined
as 
\begin{equation}
\{F,G\}=\int d^4x\{\frac{\delta F}{\delta A_\mu ^a(x)}\frac{\delta G}{\delta
E^{a\mu }(x)}-\frac{\delta F}{\delta E_\mu ^a(x)}\frac{\delta G}{\delta
A^{a\mu }(x)}\}
\end{equation}
The Poisson brackets in Eqs.(17) and (18) are easily calculated. The results
are 
\begin{equation}
\{C^a(x),\varphi ^b(y)\}=D_\mu ^{ab}(x)\partial _x^\mu \delta ^4(x-y)
\end{equation}
\begin{equation}
\{\varphi ^a(x),\varphi ^b(y)\}=0
\end{equation}
\begin{equation}
\{C^a(x),C^b(y)\}=m^2[gf^{abc}A_0^c(x)-2\delta ^{ab}\partial _0^x]\delta
^4(x-y)
\end{equation}
\begin{equation}
\{H,\varphi ^a(x)\}=\partial _x^kE_k^a(x)
\end{equation}
\begin{equation}
\{H,C^a(x)\}=m^2[\partial _0^xA_0^a(x)+\partial _k^xA_k^a(x)]
\end{equation}
It is pointed out that by the requirement of Lorentz-covariance, in the
computation of the above brackets, the second term in Eq.(10) has been
written in a Lorentz-covariant form $gf^{abc}A^{b\mu }E_\mu ^c$. We are
allowed to do it because the added term $gf^{abc}A_0^bE_0^c$ only gives a
vanishing contribution to the term $A_0^aC^a$ in Eq.(9) due to the identity $%
f^{abc}A_0^aA_0^b=0$. Particularly, the determinant of the matrix which is
constructed by the Poisson bracket denoted in Eq.(20) is not singular. This
indicates that Eqs.(17) and (18) are sufficient to determine the Lagrange
multipliers $A_0^a(x)$ and $E_0^a(x)$ respectively. There is no necessity of
taking other subsidiary constraint conditions into account further. On
substituting Eqs.(20)-(24) into Eqs.(17) and (18), we find 
\begin{equation}
\Box _xA_0^a(x)-gf^{abc}\partial _x^\mu [A_0^b(x)A_\mu ^c(x)]-\partial
_x^kE_k^a(x)=0
\end{equation}
\begin{equation}
\lbrack \delta ^{ab}\Box _x-gf^{abc}A_\mu ^c(x)\partial _x^\mu ]E_0^b(x)=0
\end{equation}
These equations are compatible with the equations (3) and (12)-(14). In
fact, as easily verified, Eqs.(25) and (26) can directly be derived from
Eqs.(3) and (12)-(14). In addition, we mention that if the Hamiltonian is
defined by $\bar {{\cal H}}={\cal H}-A_0^aC^a+E_0^a\varphi ^a$, the
equations (25) and (26) can also be obtained from the equations $\{\bar H%
,\varphi ^a(x)\}=0$ and $\{\bar H,C^a(x)\}=0$ respectively.

Now let us turn to formulate the quantization performed in the Hamiltonian
path-integral formalism for the massive non-Abelian gauge field theory. In
accordance with the general procedure of the quantization[18-20], we firstly
write the generating functional of Green's functions via the independent
canonical variables $A_T^{a\mu }$ and $E_T^{a\mu }$. 
\begin{equation}
Z[J]=\frac 1N\int D(A_T^{a\mu },E_T^{a\mu })exp\{i\int d^4x[E_T^{a\mu }\dot A%
_{T\mu }^a-{\cal H}^{*}(A_T^{a\mu },E_T^{a\mu })+J_T^{a\mu }A_{T\mu }^a]\}
\end{equation}
where ${\cal H}^{*}(A_T^{a\mu },E_T^{a\mu })$ is the Hamiltonian which is
obtained from the Hamiltonian (11) by replacing the constrained variables $%
A_L^{a\mu }$ and $E_L^{a\mu }$ with the solutions of Eqs.(3) and (14). As
mentioned before, Eq.(3) leads to $A_L^{a\mu }=0$. If setting $E_L^{a\mu
}(x)=\partial _x^\mu Q^a(x)$ where $Q^a(x)$ is a scalar function, one may
get from Eq.(14) an equation obeyed by the scalar function 
\begin{equation}
K^{ab}(x)Q^b(x)=w^a(x)
\end{equation}
where 
\begin{equation}
K^{ab}(x)=\delta ^{ab}\Box _x-gf^{abc}A_T^{c\mu }(x)\partial _\mu ^x
\end{equation}
and 
\begin{equation}
w^a(x)=gf^{abc}E_T^{b\mu }(x)A_{T\mu }^c(x)-m^2A_T^{a0}(x)
\end{equation}
With the aid of the Green's function $G^{ab}(x-y)$ (the ghost particle
propagator) which satisfies the following equation. 
\begin{equation}
K^{ac}(x)G^{cb}(x-y)=\delta ^{ab}\delta ^4(x-y)
\end{equation}
one may find the solution to the equation (28) as follows 
\begin{equation}
Q^a(x)=\int d^4yG^{ab}(x-y)w^b(y)
\end{equation}
In order to express the generating functional in terms of the variables $%
A_\mu ^a$ and $E_\mu ^a$, it is necessary to insert the following $\delta $%
-functional into the generating functional in Eq.(27)[18-20] 
\begin{equation}
\delta [A_L^{a\mu }]\delta [E_L^{a\mu }-E_L^{a\mu }(A_T^{a\mu },E_T^{a\mu
})]=detM\delta [C^a]\delta [\varphi ^a]
\end{equation}
where $M$ is the matrix whose elements are 
\begin{equation}
M^{ab}(x,y)=\{C^a(x),\varphi ^b(y)\}
\end{equation}
which was given in Eq.(20). The relation in Eq.(33) is easily derived from
Eqs.(3) and (14) by applying the property of $\delta $-functional. Upon
inserting Eq.(33) into Eq.(27) and utilizing the representation 
\begin{equation}
\delta [C^a]=\int D(\eta ^a/2\pi )e^{i\int d^4x\eta ^aC^a}
\end{equation}
we have 
\begin{eqnarray}
Z[J] &=&\frac 1N\int D(A_\mu ^a,E_\mu ^a,\eta ^a)detM\delta [\partial ^\mu
A_\mu ^a]\times exp\{i\int d^4x[E^{a\mu }\dot A_\mu ^a  \nonumber \\
&&+\eta ^aC^a-{\cal H}(A_\mu ^a,E_\mu ^a)+J^{a\mu }A_\mu ^a]\}
\end{eqnarray}
In the above exponential, there is a $E_0^a$-related term $E_0^a(\partial
_0A_0^a-\partial _0\eta ^a)$ which permits us to perform the integration
over $E_0^a$, giving a $\delta $-functional $\delta [\partial
_0A_0^a-\partial _0\eta ^a]=det|\partial _0|^{-1}\delta [A_0^a-\eta ^a]$.
The determinant $det|\partial _0|^{-1}$, as a constant, may be put in the
normalization constant $N$ and the $\delta $-functional $\delta [A_0^a-\eta
^a]$ will disappear when the integration over $\eta ^a$ is carried out. The
integral over $E_k^a$ is of Gaussian-type and hence easily calculated. After
these manipulations, we arrive at 
\begin{eqnarray}
Z[J] &=&\frac 1N\int D(A_\mu ^a)detM\delta [\partial ^\mu A_\mu
^a]exp\{i\int d^4x[-\frac 14F^{a\mu \nu }F_{\mu \nu }^a  \nonumber \\
&&+\frac 12m^2A^{a\mu }A_\mu ^a+J^{a\mu }A_\mu ^a]\}
\end{eqnarray}
When employing the familiar expression[18,19] 
\begin{equation}
detM=\int D(\bar C^a,C^a)e^{i\int d^4xd^4y\bar C^a(x)M^{ab}(x,y)C^b(y)}
\end{equation}
where $\bar C^a(x)$ and $C^a(x)$ are the ghost field variables and the
following limit for the Fresnel functional 
\begin{equation}
\delta [\partial ^\mu A_\mu ^a]=\lim_{\alpha \to 0}C[\alpha ]e^{-\frac i{%
2\alpha }\int d^4x(\partial ^\mu A_\mu ^a)^2}
\end{equation}
where $C[\alpha ]\sim \prod_x(\frac i{2\pi \alpha })^{1/2}$ and
supplementing the external source terms for the ghost fields, the generating
functional in Eq.(37) is finally given in the form 
\begin{equation}
Z[J,\bar K,K]=\frac 1N\int D(A_\mu ^a,\bar C^a,C^a)exp\{i\int d^4x[{\cal L}%
_{eff}+J^{a\mu }A_\mu ^a+\bar K^aC^a+\bar C^aK^a]\}
\end{equation}
where 
\begin{equation}
{\cal L}_{eff}=-\frac 14F^{a\mu \nu }F_{\mu \nu }^a+\frac 12m^2A^{a\mu
}A_\mu ^a-\frac 1{2\alpha }(\partial ^\mu A_\mu ^a)^2-\partial ^\mu \bar C%
^aD_\mu ^{ab}C^b
\end{equation}
which is the effective Lagrangian for the quantized massive gauge field. In
Eq.(40), the limit $\alpha \to 0$ is always implied. Certainly, the theory
may be given in general gauges$(\alpha \ne 0)$. In this case, the ghost
particle will acquire a mass $\mu =\sqrt{\alpha }m$ (we leave the details of
this subject in other papers). However, we note that the Landau gauge is
truly physical gauge for the massive gauge field which we need to work in
only for practical calculations. In passing, we note that in the zero-mass
limit, Eqs.(40) and (41) immediately go over to the corresponding results
for the massless gauge field theory established in the Lorentz gauge
condition.

From the generating functional (40), one may derive the massive gauge boson
propagator like this [6,18] 
\begin{equation}
iD_{\mu \nu }^{ab}(k)=-i\delta ^{ab}\frac{g_{\mu \nu }-k_\mu k_\nu /k^2}{%
k^2-m^2+i\varepsilon }
\end{equation}
which, as we see, can not make trouble with the renormalizability of the
theory. The ghost particle propagator and vertices derived from the above
generating functional are the same as those given in the massless theory.
Furthermore, the BRST transformation [21] under which the effective action
appearing in Eq.(40) is invariant and the Ward-Takahashi identity [6,22]
obeyed by the generating functional are formally identical to those for the
massless theory. Therefore, by the reasoning much similar to that given in
the massless theory, the renormalizability and unitarity of the theory can
be exactly proved to be no problems. All these issues will be discussed in
detail in the separate papers. We end this letter with an emphasis that the
massive gauge field theory can, indeed, be set up on the gauge-invariance
principle without relying on the Higgs mechanism. The key point to achieve
this conclusion is that the massive gauge field must be viewed as a
constrained system and the Lorentz condition must be introduced from the
beginning and imposed on the Proca Lagrangian.

\begin{center}
{\bf ACKNOWLEDGEMENT}
\end{center}

The author would like to thank professor Shi-Shu Wu for useful discussions.
This subject was supported in part by National Natural Science Foundation of
China.

\end{document}